# Dosimetric Evaluation of a New Rotating Gamma System for Stereotactic Radiosurgery

Huan Liu[1], Ahmed Eldib[1], Lili Chen[1], Bin Wang[2], Shidong Li[2], Curtis Miyamoto[2] and CM Charlie Ma[1†]

1 *Department of Radiation Oncology, Fox Chase Cancer Center, Philadelphia, PA 19111*

2 *Department of Radiation Oncology, Temple University Hospital, Philadelphia, PA 19140*

[†]Corresponding Author: Charlie.Ma@fccc.edu

**Abstract**

Purpose: A novel rotating gamma stereotactic radiosurgery (SRS) system (Galaxy® RTi) with real-time image guidance technology has been developed for high-precision SRS and frameless fractionated stereotactic radiotherapy (SRT). This work investigated the dosimetric quality of Galaxy by comparing both the machine treatment parameters and plan dosimetry parameters with those of the widely used Leksell Gamma Knife (LGK) systems for SRS.

Methods: The Galaxy RTi system uses 30 cobalt-60 sources on a rotating gantry to deliver non-coplanar, non-overlapping arcs simultaneously while the LGK 4C uses 201 static cobalt-60 sources to deliver noncoplanar beams. Ten brain cancer patients were unarchived from our clinical database, which were previously treated on the LGK 4C. The lesion volume for these cases varied from 0.1 $cm^3$ to 15.4 $cm^3$. Galaxy plans were generated using the Prowess TPS (Prowess, Concord, CA) with the same dose constraints and optimization parameters. Treatment quality metrics such as target coverage (%volume receiving the prescription dose), conformity index (CI), cone size, shots number, beam-on time were compared together with DVH curves and dose distributions.

Results: Superior treatment plans were generated for the Galaxy system that met our clinical acceptance criteria. For the 10 patients investigated, the mean CI and dose coverage for Galaxy was 1.77±0.22 and 99.24±0.33% compared to 1.94±0.48 and 99.19±0.67% for LGK, respectively. The beam-on time for Galaxy was 17.42±5.79 minutes compared to 21.34±7.35 minutes for LGK (both assuming dose rates at the initial installation). The dose fall-off is much faster for Galaxy, compared with LGK.

Conclusion: The Galaxy RTi system can provide dose distributions with similar quality to that of LGK with less beam-on time and faster dose fall-off. The system is also capable of real-time image guidance at treatment position to ensure accurate dose delivery for SRS.

# 1. Introduction

In stereotactic radiosurgery (SRS), a high dose of radiation is given to a target with a set of radiation beams converging on the target from various angles, while the surrounding healthy tissues only receive the minimal dose since a single beam is too weak to deposit enough energy. Ideally, this dose distribution results in destruction of the tumor while sparing the functions of crucial organs or tissues adjacent to the treatment area, such as the optic nerve or brainstem. The SRS procedure has evolved to allow treatment for both intracranial and extracranial sites. The most widely accepted use for SRS is still for intracranial disease. SRS is now a standard option for many malignant and benign lesions of the brain, as well as some functional conditions.

Currently, there are several radiosurgery units available. Arguably the best known SRS unit is the Gamma Knife manufactured by Elekta (Elekta AB, Stockholm, Sweden). Gamma Knife radiosurgery delivers high precision radiation treatment using multi-narrow Co-60 beams focused on a single point in a three-dimensional space. Because the radiation is heavily concentrated in the vicinity of this point, resulting in a highly peaked, near-spherical-shaped dose distribution (called a shot), multiple shots are needed to form a global radiation dose distribution to match the irregular shape of the tumor volume. The Leksell Gamma Knife (LGK) has six versions: models U, B, C, 4C, Perfexion® and ICON®. Although the physical appearance of the models differs, the internal design leads to dose profiles that vary only slightly. Each LGK system consists of the basic components: the radiation unit, the beam focusing technology, the patient couch, an electric bed system, the control console, and the planning computer system. In the C and 4C units, 201 Cobalt-60 sources were used with 4 removable tungsten circular collimators of diameters 4, 8, 14, and 18 mm to produce four different shot-sizes. An aluminum frame is attached to the patient head with four screw pins to stabilize the head during the SRS procedure and serves as a point of reference for brain scans that locate the treatment target in the brain. Optimization of the target dose distribution can be achieved using different collimators with different shot positions and weights. The Perfexion and ICON units utilize an internal collimation system with 4, 8 and 16mm collimators and 192 Cobalt-60 sources, and the ICON also integrated a cone-beam computed tomography (CBCT) scanner and an infrared camera system to support the delivery of frameless fractional stereotactic radiotherapy (SRT). The patient is first scanned with the CBCT at the imaging position for target localization and then moved into the treatment positions to receive the planned shots (Leksell et al., 1983, Lindquist et al., 2008, Lunsford et al., 1989, AlDahlawi et al., 2017, Petti et al., 2021).

Gamma Knife radiosurgery is most commonly used to treat both benign and malignant brain tumors, arteriovenous malformation (AVM), trigeminal neuralgia, acoustic neuroma, and pituitary tumors. Computed tomography (CT), magnetic resonance imaging (MRI), positron emission tomography (PET), magnetoencephalography, or cerebral angiography are used for target determination, depending on the indication. Movement of the patient couch in and out of the radiation unit, and opening of the shielding door is performed with high precision. The patient's head is moved in the focus point using very precise robotics.

The CyberKnife® (Accuracy, Sunnyvale, PA) is another noteworthy SRS system for treating intracranial lesions. It uses a single 6-MV photon beam fixed to a robotic arm, which moves the beam to different positions during the course of treatment, all converging on the treatment target. The CyberKnife differs from the Gamma Knife by employing real-time x-ray images to guide treatment, and as a result, has expanded SRS to areas outside of the brain (Dieterich et al., 2011, Ma et al., 2005, Plowman et al., 1999). For intracranial treatment, the patient is fixed to the treatment table with a firm plastic mask, and the robot is guided by a series of x-ray images of the skull taken during treatment. The position of the skull is updated in real-time, and the robot adjusts the beam to account for any skull displacement. For most treatments outside the skull, the CyberKnife requires more than just bone anatomy to guide the beam position. These areas, especially lesions that move with respiration, require the placement of gold fiducial markers near or within the target. Fiducials are usually placed as a simple outpatient procedure, similar to a needle biopsy. X-ray images capture the position of these markers and guide the robot during treatment to correct for movement. For targets that move with respiration, the CyberKnife can correlate the beam position with the pattern of respiration and adjust accordingly. This feature is useful for treating tumors in the lung or liver. The inherent design of CyberKnife results in different patient treatment procedures. The fiducials, if necessary, must be implanted several days before planning images can be developed. Images are then acquired as an outpatient, with one or two visits for a CT scan and a MRI if necessary. Planning is performed while the patient is at home, without the time pressure associated with the head or body frame. The planning is typically a joint effort among the radiation oncologist, surgeon, and physicist, usually within one day. Complex cases may take several iterations, occasionally requiring two or more days. Treatments are usually delivered by a radiation therapist, with a physician present for the initiation of treatment.

Recently, a novel design rotating gamma ray system, Galaxy® RTi (Akesis, Concord, CA) with real-time image guidance technology has been developed for high precision SRS (Liu et al., 2021). This Galaxy system consists of a rotating gantry, a focusing treatment head of 30 Co-60 sources, a kV cone-beam CT system and a 3D couch. The Galaxy RTi is a high-precision intracranial SRS system with real-time and in-line CBCT + kV/kV imaging. The imaging system is on the rigid ring-gantry at the treatment plane, eliminating the need to move the patient from the imaging position to the treatment position or interrupt treatment to image. It also supports fully automated intrafractional skull tracking and corrections. The Galaxy RTi combines the proven efficacy of gamma radiation with state-of-the-art rotational technology to optimize treatment planning and delivery. The system's dynamic, rotational delivery provides more flexibility in shaping the dose distribution as opposed to traditional, fixed sector-based delivery. The compact source drawer with 30 Co-60 sources, combined with the four sizes of collimators and one blocking position, offers various possibilities to shape the dose distribution. At the control desk, after the start button is pressed, the entire treatment progresses with no further user interaction, no matter how complex the collimator shot design; there is no need to interrupt the treatment for patient imaging. This Galaxy RTi system has received 510(k) clearance from the U.S. Food and Drug Administration.

Our clinical LGK system uses 201 static cobalt-60 sources to deliver non-coplanar beams while the novel rotating Gamma system, Galaxy, uses 30 cobalt-60 sources to deliver non-coplanar, non-overlapping arcs simultaneously. The inventive source drawer approach lowers the total cost of ownership and reduces downtime during source replacement from weeks to days. The dosimetric characteristics of stationary and rotational cobalt-60 source configurations have been investigated in detail using Monte Carlo simulations (Cheung et al., 2006, Tian et al., 2016). The results showed that rotating gamma systems (modeled as having the same latitude angles, source to focus distance, and the distance from the source to the end of the collimator as the LGK static models) have smaller beam profile penumbra in the z direction, while LGK static source design has smaller penumbra in the x and y directions. The differences are more significant in using larger collimators (Cheung et al., 2006). The penumbral width of a single source is smaller for the LGK compared to the rotating gamma system, XGD (OUR United, Xian, China) due the smaller source size (Gamma Knife 1mm vs. OUR XGD 2.6mm). However, the penumbral width for the combined shot distribution irradiated by all the sources is improved with the rotational source configuration with smaller x-y penumbra for collimators greater than 10mm (Tian et al., 2016).

This work investigated the treatment plan quality of the Galaxy RTi by comparing both the treatment parameters and plan dosimetry parameters with those of the LGK system, which is widely used clinically for SRS of intracranial tumors.

## 2. Methods and Materials

2.1 The Galaxy RTi Rotating Gamma System

The Galaxy RTi system is an advanced gamma stereotactic radiosurgery system with continuous 360° source rotation and real-time image guidance technology (Liu et al., 2021, Goetsch et al., 1999, Ma et al., 2015, Fareed et al., 2018). This Galaxy system consists of a rotating gantry, a focusing Gamma ray treatment head, a kV cone-beam CT system and a 3D couch, as shown in Figure 1. The Galaxy RTi is the only high-precision intracranial gamma system with real-time, in-line CBCT and kV/kV imaging. As indicated in Figure 2, the imaging system, including the kV x-ray source and flat panel detector, is on the rigid ring-gantry at the treatment plane, eliminating the need to move the patient from the imaging position to the treatment position or interrupt treatment to image. The system achieves faster overall treatment times compared to other systems due to a simplified treatment workflow. It also supports fully automated intrafractional skull tracking and real-time motion corrections.

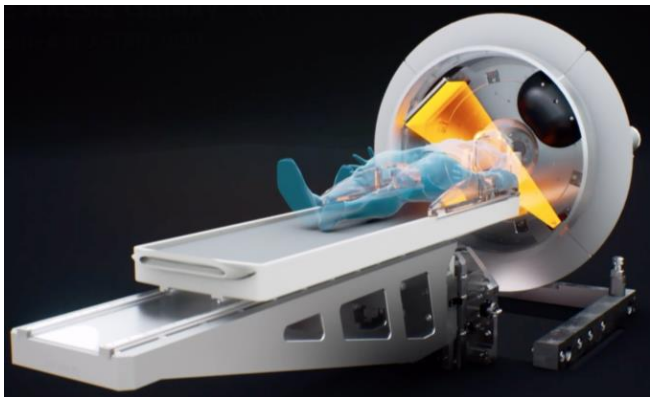

**Figure 1.** Galaxy® RTi system overview, including a rotating gantry, a focusing Gamma ray treatment head, a kV cone-beam CT system and a 3D couch.

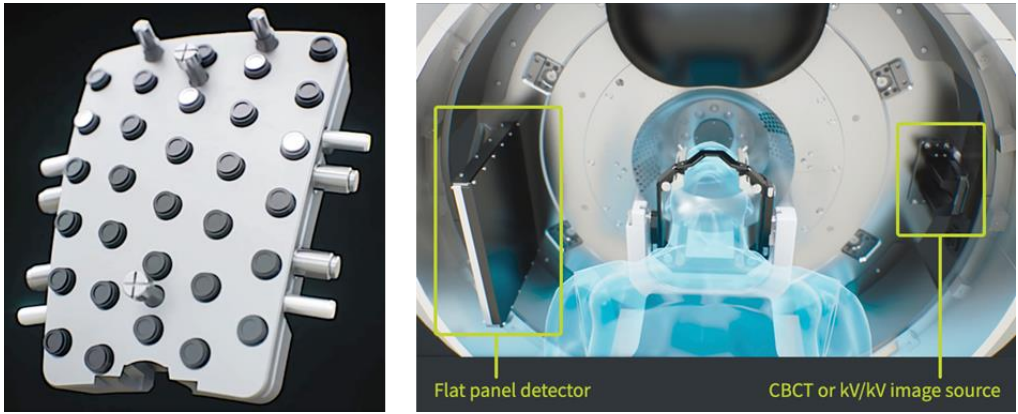

**Figure 2.** The component of the Galaxy RTi system: the compact drawer source container (left), and the real-time, in-line CBCT and kV/kV imaging system (right).

The Gamma ray head has 30 Cobalt-60 sources which are housed in a compact drawer source container, as shown in Figure 2. All sources are focused to the isocenter, and four collimator sizes (4mm, 8mm, 14mm, 18mm) and one blocking position are available. The sources can be switched off and on at any angle. The source unit and the secondary collimators rotate as a single unit. The gantry can rotate 360 degrees with a maximum speed of four rotations per minute.

The dynamic, rotational delivery of the Galaxy RTi provides more flexibility in shaping the dose distribution as opposed to traditional, fixed sector-based delivery. The compact source drawer with 30 gamma sources, combined with the four sizes of collimators (4mm, 8mm, 14mm, 18mm) and one blocking position, offers various possibilities to shape the dose distribution to match the irregular shape of the target volume.

Figure 3 shows the detailed blueprint for the Galaxy system. The source-to-axis distance (SAD) is 38.8 cm. Figure 4 shows the detailed source position in the source container. The initial dose rate of a cobalt-60 source when installed is 18.2 cGy/min (± 5%) as measured at 5mm depth in water at 38.8 cm SAD. The combined dose rate for the 30 cobalt-60 sources is 3 Gy/min as measured at the center of a spherical phantom of 16 cm diameter.

At the beginning of the radiosurgery treatment using the Galaxy system, a CBCT is acquired with the patient in the treatment position and registered to the planning image dataset. Subsequently, the couch offset is applied and the treatment proceeds with the preplanned shot positions. Real time KV images are taken at predetermined gantry angles and compared with the DRRs to monitor patient motion, while the treatment beam is on with necessary couch corrections. This real-time monitoring feature not only ensures the beam delivery accuracy by auto-correction of small patient movements during treatment but also allows the therapist to interrupt the treatment to avoid any other catastrophes. By rotating the gamma sources during treatment, 30 non-overlapping arcs are formed to produce a smooth dose distribution for each shot. Flexibility to choose partial arcs also enables the planner to avoid critical structures such as the eyes and brainstem.

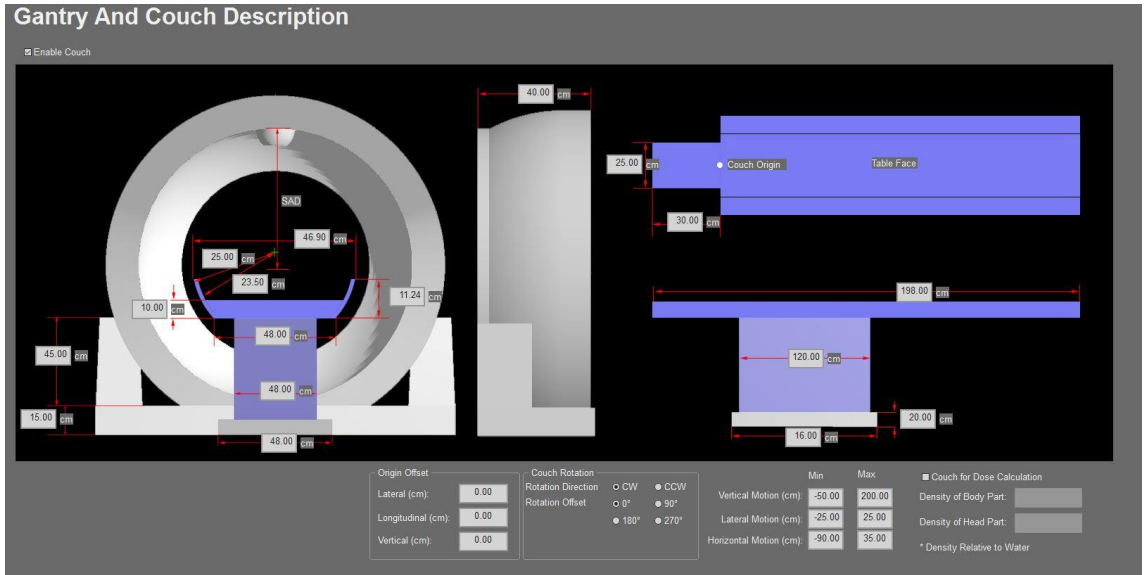

**Figure 3.** Detailed blueprint for the gantry and couch in the Galaxy system.

**Figure 4.** Source location and activity (initial dose rate).

2.2 Clinical case study and planning

Ten brain cases previously treated with the LGK 4C system at our institution between June 2018 and March 2019 were considered for this study. The imaging and plan data for those cases were unarchived in accordance with an institutional review board approval protocol. The lesion volumes ranged from 0.1 cc to 15.4 cc for various intracranial sites including brain stem, pituitary, left/right frontal, meningioma, left/right parietal, left/right temporal, left/right vestibular schwan and right parafalcine. Patient characteristics are detailed in Table 1. All of the selected cases have only one target, and multiple brain metastases were not considered for this study. All patients were treated with the LGK system and the treatment plans used for this study were all clinically approved plans and were generated using the Leksell GammaPlan (LGP) software (version 10.1.1). The prescription dose for these SRS cases ranged from 13Gy to 20Gy, all for single fraction. The detailed plan parameter information can be found in Table 1.

The Prowess Panther treatment planning system (version 1.00.4557, Prowess Inc., Concord, CA) (Fareed et al., 2018, Ma et al., 2016, Eldesoky et al., 2012, Farhood et al., 2017) was used for Galaxy RTi planning. The DICOM-RT data sets for these patients were imported from the LGP to the Prowess TPS for planning. The inverse-automated optimization (IAO) method provided by Prowess TPS was used for planning, optimizing the plans automatically by adjusting the function items such as conformity, homogeneity, dose fall-off gradient and delivery efficiency. The number of shots and collimator sizes were automatically selected by the optimization engine, depending on the shape and size of the PTV. The dose calculation was performed using a collapsed cone convolution superposition algorithm installed in the Prowess TPS (Ahnesjö et al., 1989).

Table 1 Patient information treated on the LGK system.

| Case | Site | Target Volume(cc) | Pdose (cGy) |
|---|---|---|---|
| 1 | Brain Stem | 0.5 | 1300 |
| 2 | Right Tem | 3.2 | 1300 |
| 3 | Parafalcine Pituitary | 0.1 | 1800 |
| 4 | Meningioma | 1.3 | 1300 |
| 5 | Left Parietal | 12.3 | 1800 |
| 6 | Left Front | 7.4 | 1800 |
| 7 | Right Superior | 0.7 | 2000 |
| 8 | Left Tem | 1.2 | 2000 |
| 9 | RT Frontal | 15.4 | 1800 |
| 10 | RT Vestibular Schwan | 0.8 | 1300 |

Plan quality metrics of the Galaxy plans such as target coverage (% volume receiving the prescription dose), D99% (dose received by 99% of the PTV), conformity index (CI), cone size, number of shots, beam-on time, and DVH parameters, were further evaluated and summarized.

2.3 Plan quality evaluation

The treatment plan quality of the LGK and Galaxy RTi delivery systems were compared based on the dose volume histograms (DVH), isodose distribution, PTV coverage, conformity index (CI), cone size, number of shots and beam-on time.

The CI was calculated based on the RTOG definition (Shaw et al., 1993) as a ratio of the volume covered by the prescription isodose to the target volume designated as the PTV. Typically for stereotactic radiotherapy treatment, the PTV coverage was defined as the percentage volume of PTV that received the prescription dose (Benedict et al., 2010).

For the dosimetric parameters and beam-on time, the mean quantity and standard deviation (SD) for the 10 cases were calculated, and the results were expressed as mean ± SD. Box plot of beam-on time, CI, target coverage, and D99% from both the Galaxy and LGK were used to show the distribution for those items.

The DVH curves from the LGK and Galaxy plans were exported from LGP and Prowess, respectively. All DVH curves for these 10 cases were plotted together for comparison. The dose distributions and fall-offs outside the target volumes were shown through isodose displays.

**3. Results**

3.1 Plan quality comparison for Galaxy and LGK

Superior treatment plans were generated for the Galaxy system that met our clinical acceptance criteria. Table 2 shows the prescription isodose, number of shots, cone size, beam-on time, CI, coverage and D99% from both Galaxy and LGK for the 10 patients investigated in this study. The mean CI and dose coverage for Galaxy was 1.77±0.22 and 99.24±0.33%, compared to 1.94±0.48 and 99.19±0.67% for the LGK, respectively. The beam-on time for the Galaxy system was 17.42±5.79 minutes compared to 21.34±7.35 minutes for the LGK system, both assuming dose rates at the initial installation.

The median, minimum, maximum, Q1 and Q3 of beam-on time, CI, coverage and D99% from the Galaxy and LGK systems are summarized in Table 3, whereas the boxplots for those quality comparison matrices are shown in Figure 5. Overall, the beam-on time and CI for the Galaxy system is much better than those for the LGK system with similar coverage and D99% in both treatment plans.

3.2 Further evaluation plan dose distributions

Figure 6 shows the DVH curves of the PTV for 10 patients from both the Galaxy and LGK systems. For each patient case, the solid line is the Galaxy DVH curve and the dotted line represents the LGK DVH curve. Overall, both systems produced similar dose distributions based on the prescription doses for each patient.

Table 2. The prescription isodose, number of shots, cone size, beam-on time, CI, PTV coverage and D99% from the Galaxy and LGK plans for the 10 patients investigated in this study.

| Case | System | Prescription Isodose | Shots No | Cone Size (mm) | Beam-on time (min) | CI | Coverage (%) | D99% (cGy) |
|---|---|---|---|---|---|---|---|---|
| 1 | Galaxy | 50.0 | 4 | 4, 8 | 11.6 | 1.49 | 99.4 | 1306 |
| 1 | LGK | 51.4 | 4 | 4, 8 | 17.8 | 1.40 | 98.0 | 1255 |
| 2 | Galaxy | 50.0 | 9 | 8 | 20.6 | 1.93 | 99.1 | 1302 |
| 2 | LGK | 50.0 | 14 | 8 | 30.1 | 1.70 | 99.8 | 1370 |
| 3 | Galaxy | 50.0 | 4 | 4 | 19.3 | 1.66 | 99.0 | 1868 |
| 3 | LGK | 50.0 | 4 | 4 | 29.2 | 2.80 | 98.2 | 1769 |
| 4 | Galaxy | 50.0 | 4 | 8 | 15.1 | 1.61 | 99.3 | 1321 |
| 4 | LGK | 50.0 | 5 | 8 | 20.0 | 1.40 | 99.5 | 1323 |
| 5 | Galaxy | 50.0 | 11 | 18 | 19.1 | 1.69 | 99.2 | 1814 |
| 5 | LGK | 49.8 | 11 | 18 | 26.3 | 1.54 | 99.8 | 1913 |
| 6 | Galaxy | 52.0 | 5 | 8, 14, 18 | 20.0 | 1.62 | 99.5 | 1871 |
| 6 | LGK | 44.0 | 5 | 18 | 16.1 | 1.65 | 99.0 | 1800 |
| 7 | Galaxy | 55.0 | 2 | 4, 14 | 10.1 | 1.88 | 100 | 2010 |
| 7 | LGK | 45.0 | 2 | 8, 14 | 21.3 | 2.40 | 99.2 | 2121 |
| 8 | Galaxy | 50.0 | 2 | 4, 14 | 13.6 | 2.17 | 98.9 | 1969 |
| 8 | LGK | 50.0 | 1 | 14 | 12.1 | 2.40 | 100.0 | 2725 |
| 9 | Galaxy | 54.0 | 9 | 14, 18 | 30.2 | 2.04 | 99.0 | 1805 |
| 9 | LGK | 50.7 | 9 | 18 | 30.1 | 2.09 | 99.4 | 1843 |
| 10 | Galaxy | 55.1 | 3 | 4, 8 | 14.6 | 1.63 | 99.0 | 1313 |
| 10 | LGK | 51.1 | 3 | 8, 14 | 10.5 | 2.00 | 99.0 | 1303 |

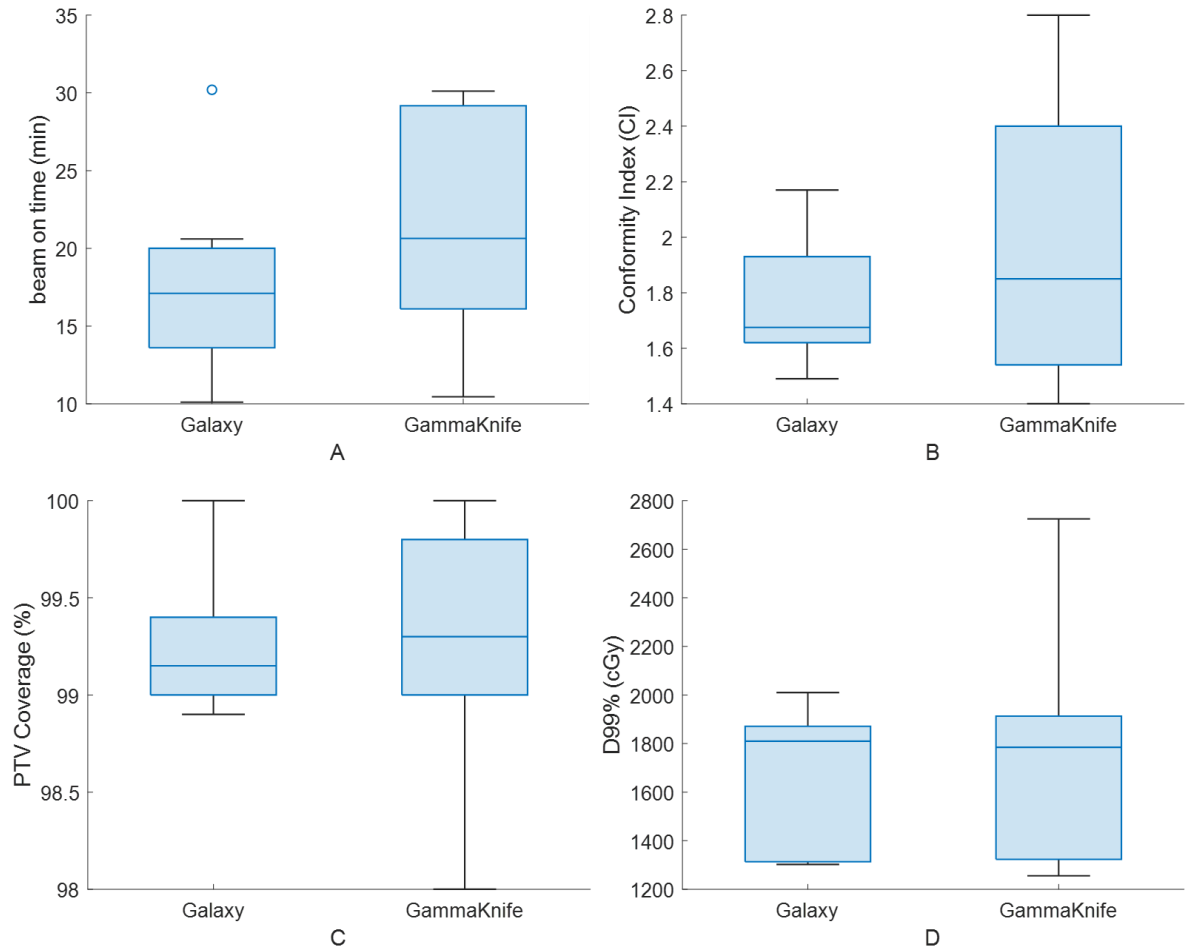

**Figure 5.** Box plots of plan quantitative metrics for both Galaxy and LGK: A is the beam-on time, B is the CI, C is the PTV coverage, and D is the D99%

**Table 3.** Results of quantitative parameters for the plans from the Galaxy and LGK systems. All values are reported as median, minimum, maximum, Q1, Q3.

| System | Beam-on time (min) | | | CI | | | Coverage (%) | | | D99% (cGy) | | |
|---|---|---|---|---|---|---|---|---|---|---|---|---|
| | median | min | Q1 | median | min | Q1 | median | min | Q1 | median | min | Q1 |
| | | max | Q3 | | max | Q3 | | max | Q3 | | max | Q3 |
| Galaxy | 17.1 | 10.1 | 13.6 | 1.68 | 1.49 | 1.62 | 99.2 | 98.9 | 99.0 | 1810 | 1302 | 1313 |
| | | 20.6 | 20.0 | | 2.17 | 1.93 | | 100.0 | 99.4 | | 2010 | 1871 |
| Gamma Knife | 20.6 | 10.5 | 16.1 | 1.85 | 1.40 | 1.54 | 99.3 | 98.0 | 99.0 | 1785 | 1255 | 1323 |
| | | 30.1 | 19.2 | | 2.80 | 2.40 | | 100.0 | 99.8 | | 2725 | 1913 |

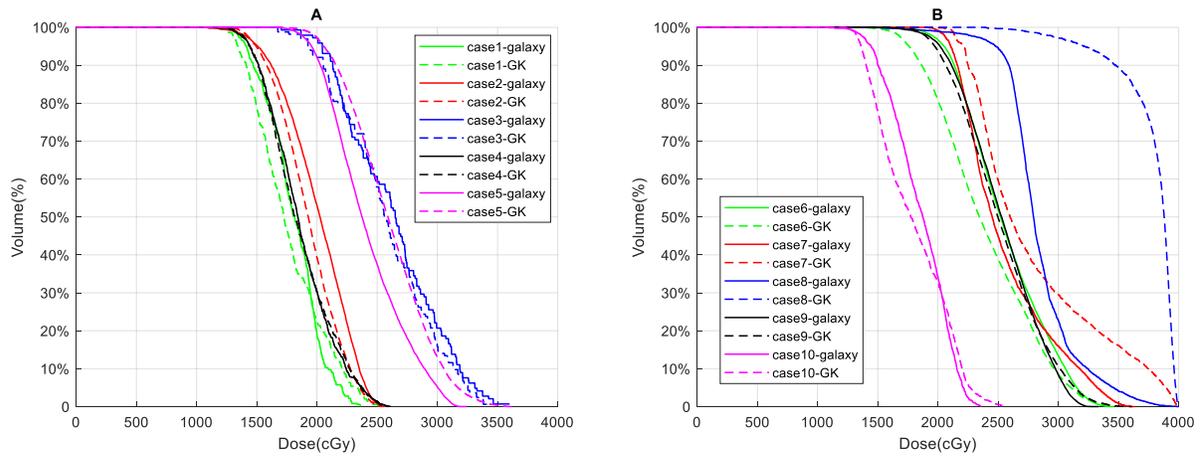

**Figure 6.** DVH comparison for Galaxy and LGK, A for case1 to case5, B for case6 to case10. Solid lines represent the DVH curves from Galaxy, and dotted lines represent the DVH curves from LGK.

Figure 7 shows the isodose line distributions for three selected patients, whose target volume was 0.5 cc (left), 3.2 cc (middle) and 15.4 cc (right), respectively. The 100% prescription isodose lines showed the same degree of conformity for the two SRS systems for the left case with a 0.5 cc target volume. However, for the other two cases with 3.2 cc and 15.4 cc target volumes, the isodose lines of 100% prescription from the Galaxy conform to the PTV much better than those from the LGK. For all three cases, the dose fall-off from the Galaxy system is much faster (with narrower distances between the isodose lines) compared with that from the LGK system.

## 4. Discussion and conclusion

The target volumes for the 10 selected patients ranged from 0.1 cc to 15.4 cc for various intracranial sites including brain stem, pituitary, left/right frontal, meningioma, etc. Compared with the LGK, the Galaxy system achieved comparative target dose coverage with slightly improved target conformity for various tumor volumes. On further comparison of the isodose distributions, the Galaxy system showed the potential to better cover the medium to large size tumor targets due to the use of advanced auto-optimization capabilities and the smooth shot dose distribution as a result of the 30 rotating cobalt sources. On the other hand, the LGK plans were generated with limited time during routine Gamma Knife SRS treatment procedures, which may be further improved with finer adjustments of shots positions and weights. For all tumor sizes, the galaxy system produced a faster dose fall-off, which is critical for sparing organs at risk.

Generally speaking, it is accepted that SRS treatments require longer delivery times than conventional radiotherapy because of the higher fractional dose and delivery accuracy. However, it is still desirable to reduce the SRS treatment time as much as possible to minimize the effect of potential intrafractional motion. The beam-on time of the plans using the Galaxy system was 17.42±5.79 minutes for the 10 patients investigated, which is significantly reduced compared with the LGK plans, where the beam-on time was 21.34±7.35 minutes. Often, a few minutes can make a difference in helping patients in poor health.

The Galaxy system has only 30 sources, but can achieve comparative PTV dose coverage. The inventive source drawer with a small number of cobalt sources can significantly reduce the cost of source replacement and routine machine maintenance, which is vital for people from developing countries to gain access to high quality cancer care.

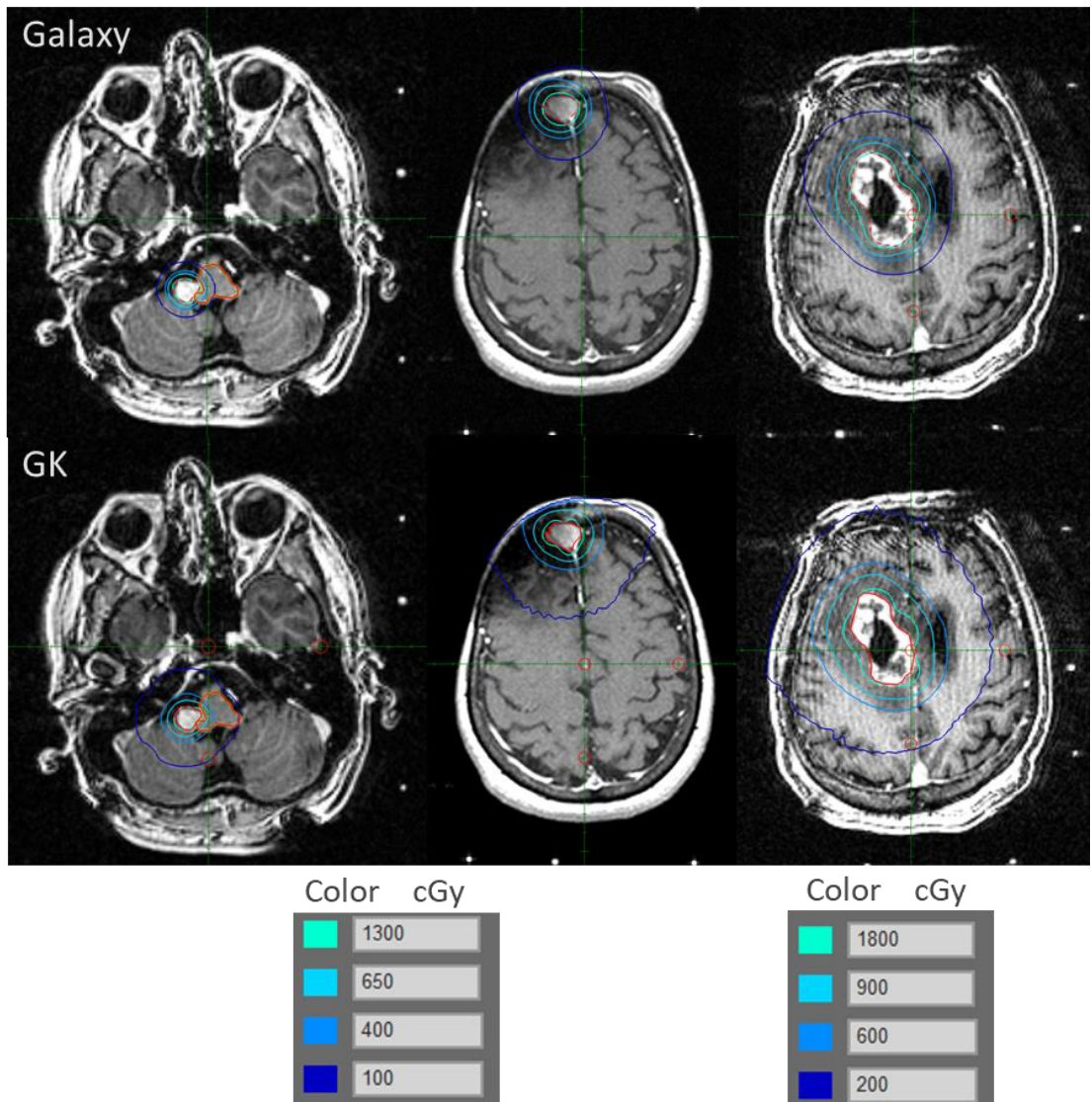

**Figure 7.** Isodose line distributions for three cases. The results from the Galaxy system are in the top row, and the results from the LGK are in the bottom row. The red dotted line is the contour of PTV for Galaxy; the red solid line is the contour of PTV for LGK. The left and middle cases had the same prescription dose of 13Gy, and the isodose lines plotted are 13Gy, 6.5Gy, 4Gy and 1Gy. The prescription dose for the right case was 18Gy, and the isodose lines plotted are 18Gy, 9Gy, 6Gy and 2Gy.

In conclusion, the Galaxy RTi system can provide dose distributions with a similar quality to that of LGK system with less beam-on time and faster dose fall-off. The system is also capable of real-time image guidance at treatment position to ensure accurate dose delivery for SRS.

**Conflict of interest**

The authors declare that they have no conflict of interest.